\documentclass[twocolumn,showpacs,preprintnumbers,amsmath,amssymb,aps,pre,floatfix,superscriptaddress,a4paper]{revtex4}
\usepackage{graphicx}
\usepackage{textcomp}
\usepackage{times}
\usepackage{amssymb}
\usepackage{txfonts}
\usepackage{amsmath}
\usepackage{color}
\usepackage{ulem}
\usepackage{gensymb}
\graphicspath{{.}}%
\newcommand{\mylab}[1]{\label{#1}}
\begin{document}
\title{Spontaneous polarization in an interfacial growth model for
  actin filament networks with a rigorous mechano-chemical coupling}
\author{Karin John}
\email{karin.john@ujf-grenoble.fr}
\affiliation{Univ. Grenoble Alpes, LIPHY, F-38000 Grenoble, France}
\affiliation{CNRS, LIPHY, F-38000 Grenoble, France}
\author{Denis Caillerie}
\affiliation{Univ. Grenoble Alpes, 3SR, F-38000 Grenoble, France}
\affiliation{CNRS, 3SR, F-38000 Grenoble, France}
\author{Chaouqi Misbah}
\email{chaouqi.misbah@ujf-grenoble.fr}
\affiliation{Univ. Grenoble Alpes, LIPHY, F-38000 Grenoble, France}
\affiliation{CNRS, LIPHY, F-38000 Grenoble, France}
\begin{abstract}
  Many processes in eukaryotic cells, including cell motility, rely on
  the growth of branched actin networks from surfaces. Despite its
  central role the mechano-chemical coupling mechanisms which guide
  the growth process are poorly understood, and a general continuum
  description combining growth and mechanics is lacking. We develop a
  theory that bridges the gap between mesoscale and continuum limit
  and propose a general framework providing the evolution law of actin
  networks growing under stress. This formulation opens an area
  for the systematic study of actin dynamics in arbitrary geometries.
%
%
%
Our framework predicts a morphological instability of actin growth on a rigid sphere,  leading to a spontaneous
polarization of the network with a mode selection corresponding to a comet, as reported experimentally.
We show that the mechanics of the contact between the
network and the surface plays a
crucial role, in that it determines directly the existence of the instability. We extract scaling laws relating growth dynamics
and network properties offering basic perspectives for new experiments on growing actin networks.
\end{abstract}
\maketitle
%
%
\section{Introduction}
Cells often migrate in response to external signals, including
chemical and mechanical signals. Thereby
the interfacial growth of filamentous actin polymer networks plays an
important role \cite{SvBo99,LBP99}. For example, cell crawling on a
two-dimensional substrate involves
the formation of a cytoplasmic membrane protrusion pointing in the direction of motion.
Thereby, the necessary force for extending the membrane is provided by
the polymerization of actin, a process far
from chemical equilibrium, which converts chemical into
mechanical energy. 
The same molecular machinery is also responsible for
the propulsion of cellular organelles \cite{TRM00}, pathogens
\cite{YTA99,CCG95} or biomimetic objects, such as spherical beads \cite{OuT99,NGF00,Bernheim_Science_2002,GPP05,DSR08,AMMG10},
vesicles \cite{UCA03,GFT03,DSR08}, droplets
\cite{BCJ04}, and ellipsoids \cite{LSZT12}.
In contrast, cell motion in a three-dimensional
substrate is rather driven by blebbing, which relies on the contraction of
the actin cortex by myosin motors to form protrusions \cite{Charras_NatRevMolCellBiol_2008}.

On the time scale of actin filament growth ($\sim 1\,$s) actin
networks behave as nonlinear elastic solids
\cite{GSMM04,JMRL07}. Typically, the linear elastic modulus of actin
networks formed during the propulsion of pathogens or biomimetic
objects is in the range of 1 to 10\,kPa \cite{GCR00,PRFH12}.
This
raises the question of the coupling mechanism between growth dynamics
and deformations (or stresses) in the network, a property which is
often either neglected \cite{DALR09,OSS08,Kawska_PNAS_2012} or included via ad hoc
assumptions, which do
not necessarily respect all symmetries in the system \cite{SPJ04,GPP05}.
Our objective is therefore
to derive macroscopic evolution equations of actin networks combining
a macroscopic constitutive law for its mechanical behavior
and the actin polymerization kinetics in a rigorous way.
The general formulation we present can be adapted to
  diverse situations relevant for actin dynamics. As a proof of
  principle we treat here the case of an elastic actin network growing from
  a spherical surface, where a spontaneous symmetry-breaking and the
  onset of motion has been
  observed experimentally
  \cite{OuT99,NGF00,Bernheim_Science_2002,GPP05,DSR08,AMMG10}. 

To briefly outline the experimental observations (more details can be
found in \cite{Blanchoin_PhysRev_2014}), actin polymerizes on the surface of a sphere with a radius of
about 1\,$\mu$m into a cross-linked/\-entangled network,
which forms initially a closed spherical shell.
Growth and cross-linking is restricted to a zone close to the
surface of the sphere.
Monomers diffuse freely through the network to reach the growth
zone and are inserted between the already existing network and the
sphere, which leads to a relative motion between older network layers
and the surface and to the buildup of mechanical stresses. Eventually
the actin shell undergoes a spontaneous symmetry-breaking, leading to the formation of an actin
tail. Two mechanisms for symmetry-breaking have been proposed:
 (i) the external actin shell ruptures at one point due to
 elongational stresses \cite{GPP05},
 or (ii) the instability is due to actin polymerizing slower on one
 side of the surface than on the other side \cite{DSR08}.

Our understanding of these processes has been advanced through
several theoretical studies based on discrete models on the scale of the filament
\cite{OuT99,MoO03}, mesoscopic models \cite{DALR09,OSS08,Kawska_PNAS_2012} or
phenomenological continuous models
\cite{SPJ04,LLK05,GPP05,JPK08}.
Rather surprisingly,  a general growth law based on first
thermodynamic principles which links the stresses in
the network (which depend on the growth history itself and the
relevant boundary conditions) to the interface dynamics is lacking. Our approach closes this gap. In this brief exposition, we exploit the framework  to show that an instability arises from the interplay between interfacial growth
kinetics and mechanical stresses, which reflect the growth history of
the network.
Thereby the nature of the contact formed between the network and the
bead surface (fixed vs. sliding) is crucial in the overall macroscopic
dynamics. In a spherical geometry a spontaneous
polarization into front and back is possible when the filaments can
slide on the surface, but not when they are fixed to the surface.
Furthermore we derive scaling laws for the instability characteristics
which form a consistent picture with experiments.

The problem of symmetry-breaking has been studied theoretically on the continuum
  level in \cite{SPJ04,JPK08}. However, in contrast to
  \cite{SPJ04,JPK08} we describe here the buildup of stresses and
  strains in the network due to growth and the subsequent
  mechano-chemical coupling in a rigorous way, consistent with a
  hyperelastic macroscopic constitutive law of the network. For
  example, the model in \cite{SPJ04} tacitly assumes a vanishing
  Poisson ratio and the stress distribution can only be solved
  consistently in an axisymmetric or spherical configuration. The model in \cite{JPK08} is limited to small deformations
  arising from a small displacment at the internal bead/network
  interface. Moreover,
  our description distinguishes between reference and deformed network
  configurations, which is essential for describing a growing interface
  in contact with a solid stationary substrate (the spherical surface), where the
  growth process manifests itself as a displacement of a free interface
  (in contact with the solvent). This crucial distinction
  between reference and deformed frame is absent
  in \cite{SPJ04,JPK08} and limits their predictive power.

\section{Model}{
\subsection{Mechanical description of the  network}
Recently we have proposed a macroscopic mechanical model of actin
networks starting from a microscopic description and have shown that it
captures the basic bulk rheological properties of actin networks
\cite{John_PhysRevE2013}.
A major issue is now, how to properly combine interfacial growth and
mechanics. We first briefly recall the description of the network
mechanics and then introduce the dynamical equations for the
interfaces.  For simplicity, we consider a 2D geometry (albeit a 3D
study does not pose a specific challenge, but increases the technical
complexity).

We assume a structurally periodic planar filament
network with the topology shown in Fig. \ref{network:topo}, which is
in contact with the surface of a cylinder of radius $R_0$.
The typical scale of the elementary cell $l_0$ is small compared to a
macroscopic scale $R_0$, which introduces
the small parameter $\eta=l_0/R_0\ll 1$.
\begin{figure}
\includegraphics[width=0.7\hsize]{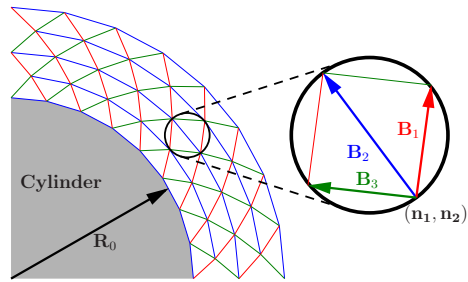}
\caption{Sketch of a small part of the filament network
  showing the different types of elementary vectors $\mathbf{B}_i$ of one
  elementary cell labeled $(n_1,n_2)$ in different colors.\mylab{network:topo}}
\end{figure}
Each elementary cell contains one node, identified by a doublet of
integers $(n_1,n_2)$, and three filaments, described by the vectors
$\mathbf{B}_i$ with $i=1,2,3$. It has a quadrilateral
  shape with the filaments $\mathbf{B}_1$ and $\mathbf{B}_3$ forming the sides and the
  filament $\mathbf{B}_2$ forming a diagonal. A node is formed by the
  intersection of six filaments, three of which belong to the same
  elementary cell as the node, the other three belonging to
  neighboring elementary cells. For example, the node ($n_1,n_2$)
  shown in Fig. \ref{network:topo} connects the filaments
  $\mathbf{B}_1$, $\mathbf{B}_2$, and $\mathbf{B}_3$ of the same node with
  the filaments $\mathbf{B}_1$ from node ($n_1-1,n_2$), $\mathbf{B}_2$ from
  node ($n_1,n_2-1$) and $\mathbf{B}_3$ from node
  ($n_1+1,n_2-1$). This network topology is one of the most simple
  topologies. Other structures involving several nodes per
  elementary cell are possible. However, the resulting mechanical
  equilibrium equations are much more involved, than for simple
  structures with only one node per elementary cell.

In a simple picture the filaments behave as entropic
springs \cite{GSMM04} and the elastic energy of the filament $\mathbf{B}_i$ is given by a harmonic potential
\begin{equation}
f_i={k\over 2 l_0}(l_i-l_0)^2 \mylab{harm:pot}
\end{equation}
with the equilibrium strand length $l_0$ and the actual strand length
$l_i = |\mathbf{B}_i|=\sqrt{\mathbf{B}_i\cdot\mathbf{B}_i}$.
We find it a bit more convenient to rewrite ($\ref{harm:pot}$) in a
different way which is valid as long as the actual length $l_i$ is
close to $l_0$. Indeed, in this case, since $(l_i^2-l_0^2)^2\simeq
4l_0^2 (l_i-l_0)^2$, we can write
\begin{equation}
f={k\over 8 l_0^3}\sum_i^{3}\left(l_i^2-l_0^2\right)^2. \mylab{en}
\end{equation}
Once the microscopic model is introduced, we can now write the
macroscopic equation. Owing to the fact that $\eta=l_0/R_0\ll 1$ one
introduces a set of continuous variables $(\lambda_1,\lambda_2)$
\cite{John_PhysRevE2013}, such that the node positions are
approximated by a continuous vector function $\boldsymbol{\phi}(\lambda_1,\lambda_2)$
and the filament vectors $\mathbf{B}_i$ are given by Taylor expansions of
$\boldsymbol{\phi}$ up to $\mathcal{O}(\eta)$
\begin{equation}
\mathbf{B}_{1,3}=\pm\mathbf{h}_1+{1\over 2}\mathbf{h}_2 \quad\mbox{and}\quad \mathbf{
B}_2=\mathbf{h}_2, \quad\mbox{with}\quad \mathbf{h}_i=\eta\partial_{\lambda_i}\boldsymbol{\phi}.
\mylab{bhdef}
\end{equation}
%
The surface in contact with the cylinder is called "inner interface"
as opposed to the ``outer interface'' in contact with the solvent.
%
In the axisymmetric situation
$\lambda_1\in(0,\Lambda_1)$ and $\lambda_2\in(0,2\pi)$ correspond to the radial and angular
directions, respectively. The unit vector in the radial direction is
given by $\mathbf{\hat r}=\cos{(\lambda_2)} \mathbf{\hat x}+\sin{(\lambda_2)} \mathbf{\hat y}$. Consequenlty, the vectors
  $\mathbf{h}_1$ and $\mathbf{h}_2$ point into the radial and angular
  directions, respectively.

In the continuum limit the discrete sum of (\ref{en}) over all nodes
can be converted into an integral over the continuous variables in the
domain $\Omega$ occupied by network, so that the total network energy
reads
\begin{equation}
F[\boldsymbol{\phi}]={1\over \eta^2}\int_\Omega f(\boldsymbol{\phi}) \,d\lambda_1\,d\lambda_2,\mylab{int:en}
\end{equation}
where by using (\ref{en}) and (\ref{bhdef}), $f$ can be expressed only
in terms of the continuous function $\boldsymbol{\phi}$.
Since growth occurs on a slow time scale, as compared to the
propagation of sound in the actin network, the system can be viewed at
each instant at mechanical equilibrium, which corresponds to a
minimum of $F$ with respect to a variation of $\boldsymbol{\phi}$. This
yields
$ 0=\partial_{\lambda_1}\mathbf{T}_1+\partial_{\lambda_2}\mathbf{T}_2 $
(divergence-free stress, as in classical linear elasticity). The
stress is defined as $\mathbf{T}_i=\partial f/\partial \mathbf{h}_i$ (the two
components of the vectors $\mathbf{T}_i$ define the stress
tensor). The $\mathbf{T}_i$ are explicitely given by
\begin{eqnarray}
\mathbf{T}_1 & = & {k\over l_0^3}\left[ (l_1^2-l_0^2) \mathbf{B}_1-
(l_3^2-l_0^2) \mathbf{B}_3\right] \mylab{T1}\\
\mathbf{T}_2 & = & {k\over l_0^3} \left[{(l_1^2-l_0^2)\over 2} \mathbf{B}_1+
(l_2^2-l_0^2) \mathbf{B}_2+{(l_3^2-l_0^2)\over 2} \mathbf{B}_3\right].\mylab{T2}
\end{eqnarray}
and define the constitutive law (relation between stress and strain).
Physically, the vector $\mathbf{T}_1$ ($\mathbf{T}_2$)
denotes the force exerted on a facet oriented in the direction $\mathbf{h}_2$ ($\mathbf{h}_1$) with length $|\mathbf{h}_2|$ ($|\mathbf{h}_1|$).
The associated  boundary conditions are
\begin{eqnarray}
|\boldsymbol{\phi}| & = & R_0 \quad\mbox{and}\quad \mathbf{T}\cdot\mathbf{t}  =  0 \quad\mbox{at the inner inteface}
\mylab{bc1}\\
\mathbf{T} & = & 0 \quad\mbox{at the outer interface}
\mylab{bc2}
\end{eqnarray}
where $\mathbf{T}$ and $\mathbf{t}$ denote the traction force and the tangent
vector at the interface.
%
(\ref{bc1}) is equivalent to a shear free inner interface (where
actin filament nucleators are present) which is everywhere in contact
with the cylinder surface and (\ref{bc2}) is equivalent to a force
free outer interface.

It has been shown \cite{John_PhysRevE2013}
that the potential energy per node $f$ [Eq.\,(\ref{en})] is equivalent to the
strain energy density of an isotropic
St.-Venant-Kirchhoff hyperelastic solid with  the Lam\'e coefficients
$\lambda=\mu=\sqrt{3} k/( 4 l_0)$, the Young's modulus
$Y=2k/(\sqrt{3}l_0)$ and the Poisson ratio $\sigma=1/3$.
This constitutive law results in a strain stiffening and negative normal forces under a simple
volume conserving shear, a typical behavior
observed experimentally for semiflexible networks \cite{GSMM04,JMRL07}.

\subsection{Growth kinetics and interface dynamics}
The main issue that remains to be addressed, is how one could link mechanics with actin growth dynamics in a consistent way, and what are the far-reaching consequences. For simplicity,
we consider the case that the network grows at
the inner  interface. That is, the topology of the network remains unchanged
while adding or subtracting an elementary material element at the interface. The cost in energy by adding (subtracting) a material
element will define the chemical potential difference $\Delta\mu$ that drives the interface using the following kinetics relating
the normal velocity $v_n$ to the chemical potential balance
\begin{equation}
v_n=-M\Delta\mu=-M(\Delta\mu_c+\Delta\mu_m)\,. \mylab{vn}
\end{equation}
$M$ is a positive mobility constant.
$\Delta\mu$ is composed of an attachment part
$\Delta\mu_c$ due to chemical bond formation
(we assume it to be a constant) and an elastic part
$\Delta\mu_m={\partial F\over\partial \Omega}$, which
is given by the functional
derivative of the total strain energy $F$ (\ref{int:en}) with respect to a
change in the shape of the network by respecting the boundary
conditions (\ref{bc1},\ref{bc2}).
We merely
  focus here on the main outcomes (details are in Appendix \ref{der:pot}).
%
For the inner interface 
we find
\begin{equation}
\Delta\mu_{mi}= {1\over \eta^2} \left[f-\mathbf{T}\cdot \mathbf{h}\right], \mylab{mumi}
\end{equation}
where $\mathbf{T}=\mathbf{T}_i\nu_i$ denotes the traction force at the
interface with $\nu_i$ being the $ith$-component of the unit normal
outward vector in the material frame $(\lambda_1,\lambda_2)$.
$\mathbf{h}=\mathbf{h}_i\nu_i$ is a measure for the displacement of the old
network interface due to the insertion of new material between the solid
cylinder and the soft network. All quantities are calculated at the interface position where the material
is added.
Intuitively one can interpret Eq. (\ref{mumi}) in the following
way. The energy cost for inserting a material element at the
internal boundary contains the straining of the material
element to the same state as neighboring interfacial elements and also
the work which is necessary to displace the old interface against the
traction force to make room
for the new material (the cylinder being rigid cannot be displaced).
Expression (\ref{mumi}) is the general form of the chemical potential
difference at the internal interface, which is valid for any material
whose constitutive law can be expressed in the form of Eqs. (\ref{T1})
and (\ref{T2}). 
%
%
At the force free external interface  the
elastic chemical potential is simply given
by $\mu_{me}={1 \over \eta^2} f$. 
Setting $\Delta\mu_m=\Delta\mu_{mi}$ and reporting (\ref{T1}) and (\ref{T2})
into (\ref{mumi}) leads to a nonlinear evolution equation (\ref{vn}) relating
growth speed and direction $v_n$ [Eq. (\ref{vn})] to the configuration of the network $\boldsymbol{\phi}$.
%
%
The resulting equation (\ref{vn}), together with (\ref{T1},\ref{T2}) and boundary conditions (\ref{bc1},\ref{bc2}) constitute the general framework that can be applied to any configuration and geometry. We treat here only the problem inspired by the actin comet formation on rigid beads.
\section{Results} 
Having defined the mechanical and dynamical equations in a consistent
manner, we will now explore some consequences.  We first consider the
case of an axisymmetric state of the network and analyse then its
linear stability with respect to a morphological perturbation.
\subsection{Axisymmetric network}
In the axisymmetric state the network geometry is described by the "radial layer number"
$\Lambda_1$. Recall that if in the discrete network the true number of
nodes in the radial direction is $N_1$ (with $N_1$ being a large integer), in the continuum limit the relation
$\Lambda_1= N_1 \eta=N_1l_0/R_0$ holds.
%
We look for
an axisymmetric configuration of the form $\boldsymbol{
\phi}=\phi_R \mathbf{\hat r}.$
%
Equations (\ref{en}-\ref{bc2}) are solved numerically using
continuation methods \cite{AUTO97}.

Figure \ref{hdmu} (a) shows the observable thickness $h$ of the network
$h=\phi_R(\Lambda_1)-R_0$ in the mechanical
equilibrium configuration depending on the radial layer number
$\Lambda_1$. The distinction between the observable thickness and the radial
layer number is important, since they are not necessarily related in a linear
manner.
\begin{figure}[hbt]
\includegraphics[width=0.8\hsize]{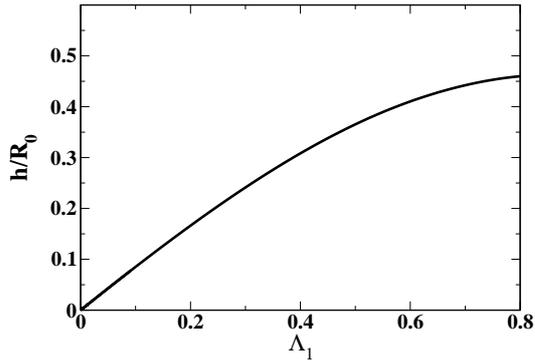}
\caption{Observable network thickness $h$ depending on the radial layer number
  $\Lambda_1$.\mylab{hdmu}}
\end{figure}
For thin networks $h$ increases linearly with
$\Lambda_1$ and one finds $h/R_0=\sqrt{3}\Lambda_1/2$. For larger
networks the tangential tension at the outer network interface leads
to a radial compression and $h$ nearly saturates for
$\Lambda_1 \simeq 0.8$.
An interesting property follows naturally from our formulation, which
is that for an axisymmetric state the elastic chemical potential
difference is identical at the two interfaces, denoted as
$\Delta\mu^{(0)}_m$.
If the thickness of the network is modified,
there is no way to discriminate
between the fact that material has been exchanged at the cylinder surface
or at the free surface, only the initial and the final thickness
matter, and not the path followed by the system.
This property is not only comforting the theory, but
can even be exploited to extract some interesting results. Indeed, it
is possible to determine analytically the chemical potential at the
external surface, which is linked to the
observable network thickness $h$ in a simple way (see Appendix
\ref{der:pot} for details)
\begin{equation}
\Delta\mu_m^{(0)}={kh^2\over 8 l_0 R_0^2}\left(h+2R_0\right)^2\,, \mylab{mum}
\end{equation}
where we have used the fact that the filaments of type 1 and 3
(cf. Fig. \ref{network:topo})
are at equilibrium length [to ensure a vanishing traction force
(\ref{bc2})] and the filaments of type 2 have length
$l_2=l_0/R_0\phi_R(\Lambda_1)=l_0(h+R_0)/R_0$.
Now we consider growth of an axisymmetric network whereby polymerization only
occurs at the internal interface with $\Delta\mu_c<0$. Dendritic actin networks nucleated by the Arp2/3 complex typically show a
kinetic polarity with a rapidly polymerizing 'plus'-end interface
directed towards the nucleating surface i.e. $\Delta\mu_{ci}<0$, and a
slowly depolymerizing 'minus'-end oriented towards the solvent,
i.e. $\Delta\mu_{ce}>0$ \cite{PBM00}. For more clarity we consider here only
the case, that the motility medium does not contain any fragmentation
proteins, such as cofilin and we treat the limiting case that
growth only occurs at the internal interface while the external
interface is stationary. The model can be straightforwardly extended
to two dynamical interfaces, which gives for example the treadmilling
behavior.

If growth occurs only at the internal interface the interface velocity
(\ref{vn}) vanishes in steady state and one finds in the
limit $h\ll 2R_0$ for the observable stationaray network thickness
\begin{equation}
\bar h  = \sqrt{-{2l_0 \Delta\mu_c \over k}}, \mylab{hss}
\end{equation}
where $\bar h$ stands for the steady thickness. Recall that in steady
state the radial layer number is related to the observable thickness
by $\bar h=\phi_R(\bar\Lambda_1)-R_0$. Expression (\ref{hss}) can be
further related to the network properties and geometry and to the
known rate equations of actin polymerization, which offers interesting
scaling relations, which will be discussed at the end of this Letter.

The next important step is to analyze the linear stability against
modulations of the gel thickness.
\subsection{Linear stability analysis}
We introduce a small perturbation of amplitude $\varepsilon$ at the
internal interface, while the structure of the external interface
remains unchanged. The perturbation is encoded in the growth velocity
at the internal surface.  
%
Since we consider a system, which is periodic (period $2\pi$) and translationally invariant in $\lambda_2$
a perturbation of any quantity (shape, strain, growth velocity) can be expressed in terms of a
$\cos$-series with wavenumber $q$ and growth rate $\beta_q$.
%
%
%
We have solved the model equations in the linear regime.
%
\begin{figure}[htb]
\includegraphics[width=0.8\hsize]{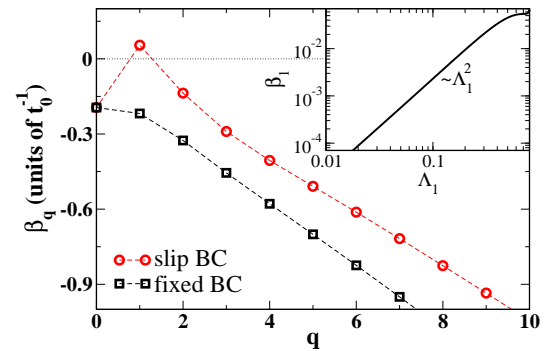}
\caption{Dispersion relation for the shear-free
  (slip BC) and the Dirichlet (fixed BC) boundary condition. Parameters are $\bar\Lambda_1=0.7$. The
  time scale is $t_0=l_0/(MkR^2_0)$. The inset shows the positive
  growth rate
  $\beta_1$ depending on the radial layer number $\Lambda_1$.\mylab{disprel}}
\end{figure}
Figure \ref{disprel} shows a typical dispersion relation. Also shown is the
dispersion relation for an alternative boundary condition, that the
network is fixed and cannot slide on the cylinder surface, i.e. condition
(\ref{bc1}) is replaced by $\boldsymbol{\phi}=R_0\mathbf{
\hat r}$ at the internal interface.
A robust and interesting outcome is that only perturbations with
the wave number $q=1$ are unstable (i.e. $\beta_1>0$), while all other
modes are stable. For relatively thin networks we find $\beta_1 =  MkR_0^2
\bar\Lambda_1^2/ l_0$. This instability is only present when the network
can slide on the cylinder surface. When the bonds are fixed, no
instability is found. The mode selection arises naturally within the
model. Damping of high modes is naturally present in the model due to the accumulation of
elastic shear-stresses in the bulk.

The instability corresponding to wave number $q=1$ means, that the
network shrinks at one side of the cylinder and grows at the opposite
side of the cylinder, i.e. the instability initiates the formation of
an actin comet as observed in biomimetic motility
experiments\cite{OuT99,GPP05,AMMG10,UCA03,GFT03,DSR08,BCJ04}.
\begin{figure}
\includegraphics[width=0.45\hsize]{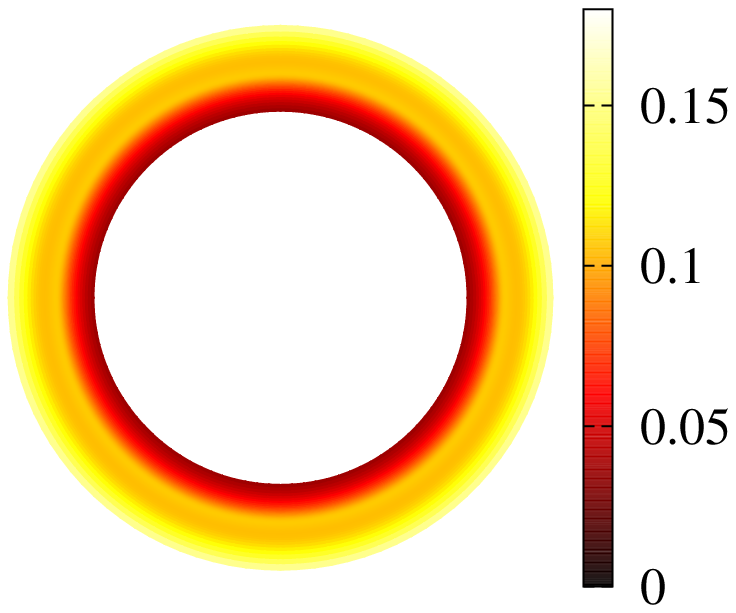} {\Large a}
\includegraphics[width=0.45\hsize]{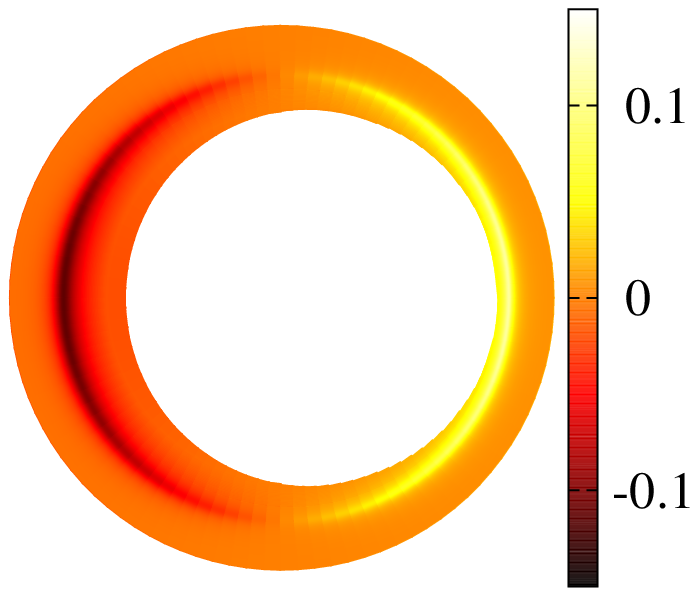}{\Large b}
\caption{Network shape before (a) and after symmetry-breaking (b). (b) was
  obtained by superimposing the axisymmetric shape and the unstable
  mode with the amplitude $0.15\,\Lambda_1$. The color encodes the
spatial distribution of the strain energy
  density (a) and the change in strain energy
  density (b). The energy density scale is $k/l_0$.\mylab{comet}}
\end{figure}
Figure \ref{comet} shows typical shapes of the network before and
after symmetry-breaking. Also shown is the strain energy density
[Fig. \ref{comet} (a)] and the change in strain energy density
compared to the axisymmetric shape [Fig. \ref{comet}
(b)]. Interestingly, the symmetry is broken by increasing the strain
energy density in the thin network regions (front) and by decreasing
it in the thick network regions (back).
As time evolves the instability should be amplified and other
mechanisms (e.g. fracture) may get important and lead
to a larger comet. A detailed study of the subsequent nonlinear regime
will be dealt with in the future.  

%
%
%
\subsection{Scaling properties and comparison with experiments}
{
In the following we will discuss some scaling relations (see Appendix \ref{scaling} for more details).
We base our calculations on the assumptions, that the
filaments behave as entropic springs with persistance length $l_p$
\cite{GSMM04}, and that free monomers (concentration $c$, size $l_m$)
polymerize into linear filaments with the known kinetics $k_+(c-c_c)$ in
the absence of mechanical stress
\cite{PBM00}, where $k_+$ denotes the rate constant of
polymerization and $c_c$ denotes the
critical monomer concentration, at which the polymerization speed is zero.
We
find the following scaling for the observable stationary network
thickness  (\ref{hss})
\begin{equation}
(\bar h/R_0)^2\sim {l_0^3(c-c_c)\over l_p^2l_m c_c}
\end{equation}
Using typical values for $l_0=100\,$nm, $l_p=10\,\mu$m, $l_m=3\,$nm
and $c/c_c=10$ we find ${\bar h/ R_0}\sim 0.2$. This value and the
linear scaling of $\bar h$ with $R_0$ are consistent with
experiments \cite{GPP05}.
The growth rate scales as $\beta\sim
k_+(c-c_c)l_m/R_0$ which yields a time scale for the birth of the
instability
\begin{equation}
\tau \sim {R_0\over l_m k_+(c-c_c)}
\end{equation}
Using $R_0=1\,\mu$m and $k_+(c-c_c)=1\,$s$^{-1}$ \cite{PBM00} one
finds a typical time $\tau=5\,$min, which is a reasonable value
\cite{GPP05}.

\section{Discussion}
We have provided a general theoretical framework for stress and actin
growth coupling. Application to actin growth on beads led to the
following major results: (i) the surprising mode selection $q=1$ and
the stabilization of all higher modes without needing an ad-hoc
cut-off length, (ii) the role of the boundary conditions at the
nucleating surface (sliding/fixed) for the existence of the
instability, (iii) the scaling laws which are consistent with
experiments. The instability reported here is induced by growth and
not by fracture. Experiments on vesicles \cite{DSR08} as a nucleating
surface are consistent with this mechanism. They suggest, that for vesicles symmetry-breaking does not occur via a
fracture mechanism at the external network interface, but via a
variation in the growth speed along the internal vesicle/network
interface. In \cite{DSR08} it was observed that shortly after 
symmetry-breaking the newly formed network layers at the internal
interface had a varying thickness (thin on one side, thick on the opposite
side of the vesicle), wereas older network layers at the external network interface
maintained a homogeneous thickness. This observation is consistent
with an asymmetric network polymerization at the internal interface.
It is
reasonable to expect bonds at the membrane to slide due to the fluid
nature of the membrane, supporting our outcome that only in this case
an instability takes place. For experiments with rigid beads
\cite{GPP05} it is not obvious (albeit not excluded) that bonds can
slide, but still an instability could be observed. It has been
proposed by the authors \cite{GPP05} that the instability is triggered
by fracture of the network. In preliminary calculations we have also
identified a morphological instability of purely mechanical origin.
However, this instability requires a critical network
thickness, beyond the thickness which we considered here. A further investigation of this problem as
well as other growth geometries (lamellipodium)  and the effect of
active stresses \cite{KJJP04} will be part of future work.

\acknowledgments C.M. acknowledges financial support from CNES (Centre d'Etudes Spatiales). The LIPhy is part of the LabEx Tec 21
(Investissements de l'Avenir - grant agreement no. ANR-11-LABX-0030).

\begin{appendix}


\section{Derivation of the elastic chemical potential difference}
\mylab{der:pot}
In the derivation of the elastic chemcical potential we will assume a
strain energy functional with the boundary conditions defined in
Eqs. (\ref{en}-\ref{bc2}) in the main text.
The elastic chemical potential difference $\Delta\mu_m$ for inserting a material
element at the network interface can be calculated from the variation
of the elastic strain energy $F$ with respect to a change in shape of the
elastic body in the material frame $\Omega$ by respecting the prescribed boundary
conditions
\begin{equation}
\Delta\mu_m={\delta F\over \delta\Omega}\,.
\end{equation}
Upon a modification of $\Omega$ by
$\delta\Omega$ the change in the strain energy is given to lowest
order in $\delta\Omega$ by
\begin{equation}
\delta F={1\over \eta^2}\int_\Omega \mathbf{T}_i \cdot\delta\mathbf{h}_i
\,d\lambda_1d\lambda_2 +{1\over \eta^2} \int_{\delta\Omega} f d\lambda_1d\lambda_2 \mylab{deltaF}
\end{equation}
with $\mathbf{T}_i=\partial f/\partial \mathbf{h}_i$ and $\mathbf{
h}_i=\eta \partial_{\lambda_i}\boldsymbol{\phi}$.
Upon partial integration of (\ref{deltaF}) one finds
\begin{eqnarray}
\delta F & = & -{1\over \eta} \int_\Omega \partial_{\lambda_i} \mathbf{T}_i\cdot\delta\boldsymbol{\phi}\,
d\lambda_1d\,\lambda_2 +{1\over \eta}\int_S \mathbf{T}\cdot\delta\boldsymbol{\phi}\,
ds + \nonumber\\
& & {1\over \eta^2} \int_{\delta\Omega} f d\lambda_1d\lambda_2 \mylab{deltaF2}
\end{eqnarray}
with $\mathbf{T}=\mathbf{T}_1\nu_1+\mathbf{T}_2\nu_2$ and where $\nu_1$ and
$\nu_2$ denote the components of the unit outward normal vector
$\boldsymbol{\nu}$ in the material frame $(\lambda_1,\lambda_2)$.

The first integral in (\ref{deltaF2}) vanishes due to the mechanical
equilibrium condition $\partial_{\lambda_i}\mathbf{T_i}=0$ and only
the second and third integral in (\ref{deltaF2}) will contribute to
the chemical potential difference.
Now, we assume that the perturbation of the interface is described by
a function
$\epsilon(|s-s_0|)\boldsymbol{\nu}(s)$, with
$\epsilon(|s-s_0|)\ll 1$.
Since the perturbation of the boundary is local and $\epsilon$ is
decaying rapidly with $|s-s_0|$ we can evaluate the third integral in (\ref{deltaF2})
to the lowest order 
\begin{equation}
\int_{\delta\Omega} f \,d\lambda_1\,d\lambda_2=\int_S f(s)\epsilon(|s-s_0|)\,ds=f(s_0)\delta\Omega\mylab{pert:prop}
\end{equation}
The second integral over the boundary of the unperturbed domain
depends crucially on the boundary condition.
First we will consider the simple case of a material exchange at the external force-free
interface with $\mathbf{T}=0$. In this case also the line integral
in (\ref{deltaF2}) vanishes and we find
\begin{equation}
\delta F={1\over \eta^2}f(s_0)\delta\Omega,
\end{equation}
which leads to the elastic chemical potential difference for a
perturbation at $s_0$ at the
external interface
\begin{equation}
\Delta\mu_{me}={1\over \eta^2}f(s_0) \mylab{dmue}
\end{equation}
In the axisymmetric configuration (\ref{dmue}) can be linked to the
observable network thickness $h=\phi_R(\Lambda_1)-R_0$ in a simple
way. Making use of the fact that the traction vector $\mathbf{T}$ at
the external interface vanishes for $l_1=l_3=l_0$ and since
$l_2=\eta\phi(\Lambda_1)=\eta(R_0+h)$, one finds after some
manipulation of (\ref{dmue}) one recovers Eq. (\ref{mum}) from the main text
\begin{equation}
\Delta\mu_{me}^{(0)}={kh^2\over 8 l_0 R_0^2}\left(h+2R_0\right)^2.
\end{equation}

Next we consider the more complex case of a material exchange at the
internal interface. Taking advantage of the boundary conditions
[Eq. (\ref{bc1}) in the main text] we note first that after a modification of the internal
boundary we find for the position of the internal boundary up to lowest order in the perturbation
\begin{equation}
R_0^2=\boldsymbol{\phi}\cdot\boldsymbol{\phi}+2\boldsymbol{\phi}\cdot(\delta\boldsymbol{\phi}+\partial_{\lambda_i}\boldsymbol{\phi}\nu_i\epsilon).
\end{equation}
The position of the  boundary before the perturbation
$\boldsymbol{\phi}$ fulfills  $|\boldsymbol\phi|=R_0$ and it follows
that
\begin{equation}
0=\boldsymbol{\phi}\cdot(\delta\boldsymbol{\phi}+\partial_{\lambda_i}\boldsymbol{\phi}\nu_i\epsilon)\quad
\mbox{and}\quad\boldsymbol{\phi}\cdot\delta\boldsymbol{\phi}=-\boldsymbol{\phi}\cdot\partial_{\lambda_i}\boldsymbol{\phi}\nu_i\epsilon.\mylab{trans}
\end{equation}
Since the vectors $\boldsymbol{\phi}$ and $\mathbf{T}$
are parallel to the normal
direction of the boundary we can write using identity (\ref{trans})
\begin{equation}
\mathbf{T}\cdot\delta\boldsymbol{\phi}={1\over R_0^2}(\mathbf{T}\cdot \boldsymbol{\phi})(\boldsymbol{\phi}\cdot\delta\boldsymbol{\phi})=-\mathbf{T}\cdot \partial_{\lambda_i}\boldsymbol{\phi}\nu_i\epsilon.
\end{equation}
After an elementary manipulation we find from (\ref{deltaF2}) and (\ref{pert:prop})
\begin{eqnarray}
\delta F & = & -{1\over \eta^2}\int_S \mathbf{T}\cdot \mathbf{h}\
\epsilon (|s-s_0|)\,ds
+{1\over \eta^2} f(s_0)\delta\Omega\nonumber\\
& = & {1\over \eta^2} \left[f-\mathbf{T}\cdot
  \mathbf{h}\right]_0\delta\Omega \mylab{dFi},
\end{eqnarray}
with $\mathbf{h}=\eta\partial_{\lambda_i}\boldsymbol{\phi}\nu_i$ and 
where the subscript $_0$ indicates that all quantities are calculated
at the position $s_0$. From (\ref{dFi}) we infer the elastic chemical
potential difference at the internal interface [Eq. (\ref{mumi}) in
the main text]
\begin{equation}
\Delta\mu_{mi}={1\over \eta^2} \left[f-\mathbf{T}\cdot\mathbf{h}\right].
\end{equation}
\section{Scaling behavior}
\mylab{scaling}
Here we demonstrate some scaling relations, relevant for biomimetic
experiments on actin driven motility.
First we note, that the polymerization part of the chemical potential $\Delta\mu_c$
used in Eq. (\ref{mumi}) in the main text is based on a material exchange in Lagrangian
coordinates. Therefore, the change in chemical potential per node is
given by $\eta^2\Delta\mu_c$, which can be related to the chemical potential difference per
monomer $\Delta\tilde\mu_c$ by
assuming that each node contains 3 filaments of length $l_0$ and that
each filament contains $l_0/l_m$ monomers
of size $l_m$. Consequently
\begin{equation}
\Delta\tilde\mu_c={l_0l_m\over
  3R_0^2}\Delta\mu_c=-k_BT\ln{c\over c_c} \mylab{muc}
\end{equation}
where $c$ denotes the actual concentration of monomers in solution, $c_c$
denotes the critical monomer concentration where the polymerization
speed vanishes and $k_BT$ denotes the thermal energy.
Using expression (\ref{muc}) we can write
\begin{equation}
\left({\bar h\over R_0}\right)^2=-{2l_0\Delta\mu_c\over k R_0^2}={6k_BT\over k l_m}\ln{c\over
  c_c}
\end{equation}
Assuming now, that the filaments behave as entropic springs with
spring constant $k=k_BTl_p^2/l_0^3$ \cite{GSMM04}, where $l_p$ denotes
the persistence length we find
\begin{equation}
\left({\bar h \over R_0}\right)^2= {6l_0^3 \over l_p^2l_m}\ln{c\over
  c_c} \approx {6l_0^3 \over l_p^2l_m c_c} (c-c_c) \mylab{hR0}
\end{equation}
Note the linear scaling of $\bar h$ with $R_0$ as observed
experimentally in \cite{GPP05}. Using typical values for
$l_0=100\,$nm, $l_p=10\,\mu$m, $l_m=3\,$nm and $c/c_c=10$ one finds
$\bar{h}/R_0\approx 0.2$.

Now, we use the known kinetic eqation of polyerization of actin to
estimate the time scale of symmetry breaking $t_0=l_0/(MkR_0^2)$,
i.e. we estimate the mobility constant $M$.
Typically the polymerization of actin without mechanical stresses is described by the following
kinetic equation \cite{PBM00}
\begin{equation}
v_p=k_+(c-c_c)\mylab{vp}
\end{equation}
where $k_+$ denotes the rate constant of polymerization. The normal interface velocity
[Eq. (\ref{vn}) in the main text] is related to the polymerization speed by $v_n\approx
l_m/R_0 v_p$ and we find
\begin{eqnarray}
v_p & = & {R_0\over l_m}v_n=-{R_0\over l_m}M\Delta\mu_c={3MR_0^3k_BT\over
  l_0l_m^2}\ln{c\over c_c} \nonumber\\
& = &{3MR_0^3k_BT\over
  l_0l_m^2c_c} (c-c_c) \mylab{vpc}
\end{eqnarray}
Comparisson of (\ref{vpc}) with (\ref{vp}) gives the following expression for the
mobility constant
\begin{equation}
M={k_+c_cl_0l_m^2 \over 3R_0^3 k_BT}
\end{equation}
and the time scale $t_0$ can be rewritten as
\begin{equation}
t_0={l_0\over M k R_0^2}={3R_0l_0^3\over k_+c_c l_p^2l_m^2} \mylab{t0}
\end{equation}
As stated in the main text, the growth rate for symmetry-breaking
scales as $\beta_1\sim \bar\Lambda^2/t_0$ for relatively thin
networks. Assuming $\bar\Lambda_1^2\sim (\bar h/R_0)^2$ and using (\ref{hR0})
and (\ref{t0}) we find the following scaling $\beta_1\sim
k_+(c-c_c)l_m/R_0$. Consequently the typical time scale for
symmetry-breaking scales linearly with $R_0$ as shown experimentally
in \cite{GPP05}. For $R_0=1\,\mu$m and $k_+(c-c_c)=1$\,$s^{-1}$
\cite{PBM00} one finds a typical time scale of symmetry-breaking of
5\,min, which is a reasonable result \cite{GPP05}.

\end{appendix}


\begin{thebibliography}{32}%
\makeatletter
\providecommand \@ifxundefined [1]{%
 \@ifx{#1\undefined}
}%
\providecommand \@ifnum [1]{%
 \ifnum #1\expandafter \@firstoftwo
 \else \expandafter \@secondoftwo
 \fi
}%
\providecommand \@ifx [1]{%
 \ifx #1\expandafter \@firstoftwo
 \else \expandafter \@secondoftwo
 \fi
}%
\providecommand \natexlab [1]{#1}%
\providecommand \enquote  [1]{``#1''}%
\providecommand \bibnamefont  [1]{#1}%
\providecommand \bibfnamefont [1]{#1}%
\providecommand \citenamefont [1]{#1}%
\providecommand \href@noop [0]{\@secondoftwo}%
\providecommand \href [0]{\begingroup \@sanitize@url \@href}%
\providecommand \@href[1]{\@@startlink{#1}\@@href}%
\providecommand \@@href[1]{\endgroup#1\@@endlink}%
\providecommand \@sanitize@url [0]{\catcode `\\12\catcode `\$12\catcode
  `\&12\catcode `\#12\catcode `\^12\catcode `\_12\catcode `\%12\relax}%
\providecommand \@@startlink[1]{}%
\providecommand \@@endlink[0]{}%
\providecommand \url  [0]{\begingroup\@sanitize@url \@url }%
\providecommand \@url [1]{\endgroup\@href {#1}{\urlprefix }}%
\providecommand \urlprefix  [0]{URL }%
\providecommand \Eprint [0]{\href }%
\providecommand \doibase [0]{http://dx.doi.org/}%
\providecommand \selectlanguage [0]{\@gobble}%
\providecommand \bibinfo  [0]{\@secondoftwo}%
\providecommand \bibfield  [0]{\@secondoftwo}%
\providecommand \translation [1]{[#1]}%
\providecommand \BibitemOpen [0]{}%
\providecommand \bibitemStop [0]{}%
\providecommand \bibitemNoStop [0]{.\EOS\space}%
\providecommand \EOS [0]{\spacefactor3000\relax}%
\providecommand \BibitemShut  [1]{\csname bibitem#1\endcsname}%
\let\auto@bib@innerbib\@empty
\bibitem [{\citenamefont {Svitkina}\ and\ \citenamefont
  {Borisy}(1999)}]{SvBo99}%
  \BibitemOpen
  \bibfield  {author} {\bibinfo {author} {\bibfnamefont {T.~M.}\ \bibnamefont
  {Svitkina}}\ and\ \bibinfo {author} {\bibfnamefont {G.~G.}\ \bibnamefont
  {Borisy}},\ }\href@noop {} {\bibfield  {journal} {\bibinfo  {journal} {J.
  Cell Biol.}\ }\textbf {\bibinfo {volume} {145}},\ \bibinfo {pages} {1009}
  (\bibinfo {year} {1999})}\BibitemShut {NoStop}%
\bibitem [{\citenamefont {Loisel}\ \emph {et~al.}(1999)\citenamefont {Loisel},
  \citenamefont {Boujemaa}, \citenamefont {Pantaloni},\ and\ \citenamefont
  {Carlier}}]{LBP99}%
  \BibitemOpen
  \bibfield  {author} {\bibinfo {author} {\bibfnamefont {T.~P.}\ \bibnamefont
  {Loisel}}, \bibinfo {author} {\bibfnamefont {R.}~\bibnamefont {Boujemaa}},
  \bibinfo {author} {\bibfnamefont {D.}~\bibnamefont {Pantaloni}}, \ and\
  \bibinfo {author} {\bibfnamefont {M.-F.}\ \bibnamefont {Carlier}},\
  }\href@noop {} {\bibfield  {journal} {\bibinfo  {journal} {Nature}\ }\textbf
  {\bibinfo {volume} {401}},\ \bibinfo {pages} {613} (\bibinfo {year}
  {1999})}\BibitemShut {NoStop}%
\bibitem [{\citenamefont {Taunton}\ \emph {et~al.}(2000)\citenamefont
  {Taunton}, \citenamefont {Rowning}, \citenamefont {Couglin}, \citenamefont
  {Wu}, \citenamefont {Moon}, \citenamefont {Mitchison},\ and\ \citenamefont
  {Larabell}}]{TRM00}%
  \BibitemOpen
  \bibfield  {author} {\bibinfo {author} {\bibfnamefont {J.}~\bibnamefont
  {Taunton}}, \bibinfo {author} {\bibfnamefont {B.~A.}\ \bibnamefont
  {Rowning}}, \bibinfo {author} {\bibfnamefont {M.~L.}\ \bibnamefont
  {Couglin}}, \bibinfo {author} {\bibfnamefont {M.}~\bibnamefont {Wu}},
  \bibinfo {author} {\bibfnamefont {R.~T.}\ \bibnamefont {Moon}}, \bibinfo
  {author} {\bibfnamefont {T.~J.}\ \bibnamefont {Mitchison}}, \ and\ \bibinfo
  {author} {\bibfnamefont {C.~A.}\ \bibnamefont {Larabell}},\ }\href@noop {}
  {\bibfield  {journal} {\bibinfo  {journal} {J. Cell Biol.}\ }\textbf
  {\bibinfo {volume} {148}},\ \bibinfo {pages} {519} (\bibinfo {year}
  {2000})}\BibitemShut {NoStop}%
\bibitem [{\citenamefont {Yarar}\ \emph {et~al.}(1999)\citenamefont {Yarar},
  \citenamefont {To}, \citenamefont {Abo},\ and\ \citenamefont
  {Welch}}]{YTA99}%
  \BibitemOpen
  \bibfield  {author} {\bibinfo {author} {\bibfnamefont {D.}~\bibnamefont
  {Yarar}}, \bibinfo {author} {\bibfnamefont {W.}~\bibnamefont {To}}, \bibinfo
  {author} {\bibfnamefont {A.}~\bibnamefont {Abo}}, \ and\ \bibinfo {author}
  {\bibfnamefont {M.~D.}\ \bibnamefont {Welch}},\ }\href@noop {} {\bibfield
  {journal} {\bibinfo  {journal} {Curr. Biol.}\ }\textbf {\bibinfo {volume}
  {9}},\ \bibinfo {pages} {555} (\bibinfo {year} {1999})}\BibitemShut {NoStop}%
\bibitem [{\citenamefont {Cudmore}\ \emph {et~al.}(1995)\citenamefont
  {Cudmore}, \citenamefont {Cossart}, \citenamefont {Griffiths},\ and\
  \citenamefont {Way}}]{CCG95}%
  \BibitemOpen
  \bibfield  {author} {\bibinfo {author} {\bibfnamefont {S.}~\bibnamefont
  {Cudmore}}, \bibinfo {author} {\bibfnamefont {P.}~\bibnamefont {Cossart}},
  \bibinfo {author} {\bibfnamefont {G.}~\bibnamefont {Griffiths}}, \ and\
  \bibinfo {author} {\bibfnamefont {M.}~\bibnamefont {Way}},\ }\href@noop {}
  {\bibfield  {journal} {\bibinfo  {journal} {Nature}\ }\textbf {\bibinfo
  {volume} {378}},\ \bibinfo {pages} {636} (\bibinfo {year}
  {1995})}\BibitemShut {NoStop}%
\bibitem [{\citenamefont {van Oudenaarden}\ and\ \citenamefont
  {Theriot}(1999)}]{OuT99}%
  \BibitemOpen
  \bibfield  {author} {\bibinfo {author} {\bibfnamefont {A.}~\bibnamefont {van
  Oudenaarden}}\ and\ \bibinfo {author} {\bibfnamefont {J.~A.}\ \bibnamefont
  {Theriot}},\ }\href@noop {} {\bibfield  {journal} {\bibinfo  {journal}
  {Nature Cell Biol.}\ }\textbf {\bibinfo {volume} {1}},\ \bibinfo {pages}
  {493} (\bibinfo {year} {1999})}\BibitemShut {NoStop}%
\bibitem [{\citenamefont {Noireaux}\ \emph {et~al.}(2000)\citenamefont
  {Noireaux}, \citenamefont {Golsteyn}, \citenamefont {Friederich},
  \citenamefont {Prost}, \citenamefont {Antony}, \citenamefont {Louvard},\ and\
  \citenamefont {Sykes}}]{NGF00}%
  \BibitemOpen
  \bibfield  {author} {\bibinfo {author} {\bibfnamefont {V.}~\bibnamefont
  {Noireaux}}, \bibinfo {author} {\bibfnamefont {R.~M.}\ \bibnamefont
  {Golsteyn}}, \bibinfo {author} {\bibfnamefont {E.}~\bibnamefont
  {Friederich}}, \bibinfo {author} {\bibfnamefont {J.}~\bibnamefont {Prost}},
  \bibinfo {author} {\bibfnamefont {C.}~\bibnamefont {Antony}}, \bibinfo
  {author} {\bibfnamefont {D.}~\bibnamefont {Louvard}}, \ and\ \bibinfo
  {author} {\bibfnamefont {C.}~\bibnamefont {Sykes}},\ }\href@noop {}
  {\bibfield  {journal} {\bibinfo  {journal} {Biophys. J.}\ }\textbf {\bibinfo
  {volume} {78}},\ \bibinfo {pages} {1643} (\bibinfo {year}
  {2000})}\BibitemShut {NoStop}%
\bibitem [{\citenamefont {Bernheim-Groswasser}\ \emph
  {et~al.}(2002)\citenamefont {Bernheim-Groswasser}, \citenamefont {Wiesner},
  \citenamefont {Golsteyn}, \citenamefont {Carlier},\ and\ \citenamefont
  {Sykes}}]{Bernheim_Science_2002}%
  \BibitemOpen
  \bibfield  {author} {\bibinfo {author} {\bibfnamefont {A.}~\bibnamefont
  {Bernheim-Groswasser}}, \bibinfo {author} {\bibfnamefont {S.}~\bibnamefont
  {Wiesner}}, \bibinfo {author} {\bibfnamefont {R.~M.}\ \bibnamefont
  {Golsteyn}}, \bibinfo {author} {\bibfnamefont {M.-F.}\ \bibnamefont
  {Carlier}}, \ and\ \bibinfo {author} {\bibfnamefont {C.}~\bibnamefont
  {Sykes}},\ }\href@noop {} {\bibfield  {journal} {\bibinfo  {journal}
  {Nature}\ }\textbf {\bibinfo {volume} {417}},\ \bibinfo {pages} {308}
  (\bibinfo {year} {2002})}\BibitemShut {NoStop}%
\bibitem [{\citenamefont {van~der Gucht}\ \emph {et~al.}(2005)\citenamefont
  {van~der Gucht}, \citenamefont {Paluch}, \citenamefont {Plastino},\ and\
  \citenamefont {Sykes}}]{GPP05}%
  \BibitemOpen
  \bibfield  {author} {\bibinfo {author} {\bibfnamefont {J.}~\bibnamefont
  {van~der Gucht}}, \bibinfo {author} {\bibfnamefont {E.}~\bibnamefont
  {Paluch}}, \bibinfo {author} {\bibfnamefont {J.}~\bibnamefont {Plastino}}, \
  and\ \bibinfo {author} {\bibfnamefont {C.}~\bibnamefont {Sykes}},\
  }\href@noop {} {\bibfield  {journal} {\bibinfo  {journal} {PNAS}\ }\textbf
  {\bibinfo {volume} {102}},\ \bibinfo {pages} {7847} (\bibinfo {year}
  {2005})}\BibitemShut {NoStop}%
\bibitem [{\citenamefont {Delatour}\ \emph {et~al.}(2008)\citenamefont
  {Delatour}, \citenamefont {Sheknar}, \citenamefont {Reyman}, \citenamefont
  {Didry}, \citenamefont {H\^o Di\^ep~L\^e}, \citenamefont {Romet-Lemonne},
  \citenamefont {Helfer},\ and\ \citenamefont {Carlier}}]{DSR08}%
  \BibitemOpen
  \bibfield  {author} {\bibinfo {author} {\bibfnamefont {V.}~\bibnamefont
  {Delatour}}, \bibinfo {author} {\bibfnamefont {S.}~\bibnamefont {Sheknar}},
  \bibinfo {author} {\bibfnamefont {A.-C.}\ \bibnamefont {Reyman}}, \bibinfo
  {author} {\bibfnamefont {D.}~\bibnamefont {Didry}}, \bibinfo {author}
  {\bibfnamefont {K.}~\bibnamefont {H\^o Di\^ep~L\^e}}, \bibinfo {author}
  {\bibfnamefont {G.}~\bibnamefont {Romet-Lemonne}}, \bibinfo {author}
  {\bibfnamefont {E.}~\bibnamefont {Helfer}}, \ and\ \bibinfo {author}
  {\bibfnamefont {M.-F.}\ \bibnamefont {Carlier}},\ }\href@noop {} {\bibfield
  {journal} {\bibinfo  {journal} {New J. Phys.}\ }\textbf {\bibinfo {volume}
  {10}},\ \bibinfo {pages} {025001} (\bibinfo {year} {2008})}\BibitemShut
  {NoStop}%
\bibitem [{\citenamefont {Achard}\ \emph {et~al.}(2010)\citenamefont {Achard},
  \citenamefont {Martiel}, \citenamefont {Michelot}, \citenamefont {Guerin},
  \citenamefont {Reymann}, \citenamefont {Blanchoin},\ and\ \citenamefont
  {Boujemaa-Paterski}}]{AMMG10}%
  \BibitemOpen
  \bibfield  {author} {\bibinfo {author} {\bibfnamefont {V.}~\bibnamefont
  {Achard}}, \bibinfo {author} {\bibfnamefont {J.-L.}\ \bibnamefont {Martiel}},
  \bibinfo {author} {\bibfnamefont {A.}~\bibnamefont {Michelot}}, \bibinfo
  {author} {\bibfnamefont {C.}~\bibnamefont {Guerin}}, \bibinfo {author}
  {\bibfnamefont {A.-C.}\ \bibnamefont {Reymann}}, \bibinfo {author}
  {\bibfnamefont {L.}~\bibnamefont {Blanchoin}}, \ and\ \bibinfo {author}
  {\bibfnamefont {R.}~\bibnamefont {Boujemaa-Paterski}},\ }\href@noop {}
  {\bibfield  {journal} {\bibinfo  {journal} {Curr. Biol.}\ }\textbf {\bibinfo
  {volume} {20}},\ \bibinfo {pages} {423} (\bibinfo {year} {2010})}\BibitemShut
  {NoStop}%
\bibitem [{\citenamefont {Upadhyaya}\ \emph {et~al.}(2003)\citenamefont
  {Upadhyaya}, \citenamefont {Chabot}, \citenamefont {Andreeva}, \citenamefont
  {Samadani},\ and\ \citenamefont {van Oudenaarden}}]{UCA03}%
  \BibitemOpen
  \bibfield  {author} {\bibinfo {author} {\bibfnamefont {A.}~\bibnamefont
  {Upadhyaya}}, \bibinfo {author} {\bibfnamefont {J.}~\bibnamefont {Chabot}},
  \bibinfo {author} {\bibfnamefont {A.}~\bibnamefont {Andreeva}}, \bibinfo
  {author} {\bibfnamefont {A.}~\bibnamefont {Samadani}}, \ and\ \bibinfo
  {author} {\bibfnamefont {A.}~\bibnamefont {van Oudenaarden}},\ }\href@noop {}
  {\bibfield  {journal} {\bibinfo  {journal} {Proc. Nat. Acad. Sci.}\ }\textbf
  {\bibinfo {volume} {100}},\ \bibinfo {pages} {4521} (\bibinfo {year}
  {2003})}\BibitemShut {NoStop}%
\bibitem [{\citenamefont {Giardini}\ \emph {et~al.}(2003)\citenamefont
  {Giardini}, \citenamefont {Fletcher},\ and\ \citenamefont {Theriot}}]{GFT03}%
  \BibitemOpen
  \bibfield  {author} {\bibinfo {author} {\bibfnamefont {P.}~\bibnamefont
  {Giardini}}, \bibinfo {author} {\bibfnamefont {D.}~\bibnamefont {Fletcher}},
  \ and\ \bibinfo {author} {\bibfnamefont {J.}~\bibnamefont {Theriot}},\
  }\href@noop {} {\bibfield  {journal} {\bibinfo  {journal} {Proc. Nat. Acad.
  Sci.}\ }\textbf {\bibinfo {volume} {100}},\ \bibinfo {pages} {6493} (\bibinfo
  {year} {2003})}\BibitemShut {NoStop}%
\bibitem [{\citenamefont {Boukellal}\ \emph {et~al.}(2004)\citenamefont
  {Boukellal}, \citenamefont {Camp\'as}, \citenamefont {Joanny}, \citenamefont
  {Prost},\ and\ \citenamefont {Sykes}}]{BCJ04}%
  \BibitemOpen
  \bibfield  {author} {\bibinfo {author} {\bibfnamefont {H.}~\bibnamefont
  {Boukellal}}, \bibinfo {author} {\bibfnamefont {O.}~\bibnamefont {Camp\'as}},
  \bibinfo {author} {\bibfnamefont {J.-F.}\ \bibnamefont {Joanny}}, \bibinfo
  {author} {\bibfnamefont {J.}~\bibnamefont {Prost}}, \ and\ \bibinfo {author}
  {\bibfnamefont {C.}~\bibnamefont {Sykes}},\ }\href@noop {} {\bibfield
  {journal} {\bibinfo  {journal} {Phys. Rev. E}\ }\textbf {\bibinfo {volume}
  {69}},\ \bibinfo {pages} {061906} (\bibinfo {year} {2004})}\BibitemShut
  {NoStop}%
\bibitem [{\citenamefont {Lacayo}\ \emph {et~al.}(2012)\citenamefont {Lacayo},
  \citenamefont {Soneral}, \citenamefont {Zhu}, \citenamefont {Tsuchida},
  \citenamefont {Footer}, \citenamefont {Frederick}, \citenamefont {Lu},
  \citenamefont {Xia}, \citenamefont {Mogilner},\ and\ \citenamefont
  {Theriot}}]{LSZT12}%
  \BibitemOpen
  \bibfield  {author} {\bibinfo {author} {\bibfnamefont {C.~I.}\ \bibnamefont
  {Lacayo}}, \bibinfo {author} {\bibfnamefont {P.~A.~G.}\ \bibnamefont
  {Soneral}}, \bibinfo {author} {\bibfnamefont {J.}~\bibnamefont {Zhu}},
  \bibinfo {author} {\bibfnamefont {M.~A.}\ \bibnamefont {Tsuchida}}, \bibinfo
  {author} {\bibfnamefont {M.~A.}\ \bibnamefont {Footer}}, \bibinfo {author}
  {\bibfnamefont {S.~S.}\ \bibnamefont {Frederick}}, \bibinfo {author}
  {\bibfnamefont {Y.}~\bibnamefont {Lu}}, \bibinfo {author} {\bibfnamefont
  {Y.}~\bibnamefont {Xia}}, \bibinfo {author} {\bibfnamefont {A.}~\bibnamefont
  {Mogilner}}, \ and\ \bibinfo {author} {\bibfnamefont {J.~A.}\ \bibnamefont
  {Theriot}},\ }\href@noop {} {\bibfield  {journal} {\bibinfo  {journal} {Mol.
  Biol. Cell}\ }\textbf {\bibinfo {volume} {23}},\ \bibinfo {pages} {615}
  (\bibinfo {year} {2012})}\BibitemShut {NoStop}%
\bibitem [{\citenamefont {Charras}\ and\ \citenamefont
  {Paluch}(2008)}]{Charras_NatRevMolCellBiol_2008}%
  \BibitemOpen
  \bibfield  {author} {\bibinfo {author} {\bibfnamefont {G.}~\bibnamefont
  {Charras}}\ and\ \bibinfo {author} {\bibfnamefont {E.}~\bibnamefont
  {Paluch}},\ }\href@noop {} {\bibfield  {journal} {\bibinfo  {journal} {Nat.
  Rev. Mol. Cell Biol.}\ }\textbf {\bibinfo {volume} {9}},\ \bibinfo {pages}
  {730} (\bibinfo {year} {2008})}\BibitemShut {NoStop}%
\bibitem [{\citenamefont {Gardel}\ \emph {et~al.}(2004)\citenamefont {Gardel},
  \citenamefont {Shin}, \citenamefont {MacKintosh}, \citenamefont {Mahadevan},
  \citenamefont {Matsudaira},\ and\ \citenamefont {Weitz}}]{GSMM04}%
  \BibitemOpen
  \bibfield  {author} {\bibinfo {author} {\bibfnamefont {M.~L.}\ \bibnamefont
  {Gardel}}, \bibinfo {author} {\bibfnamefont {J.~H.}\ \bibnamefont {Shin}},
  \bibinfo {author} {\bibfnamefont {F.~C.}\ \bibnamefont {MacKintosh}},
  \bibinfo {author} {\bibfnamefont {L.}~\bibnamefont {Mahadevan}}, \bibinfo
  {author} {\bibfnamefont {P.}~\bibnamefont {Matsudaira}}, \ and\ \bibinfo
  {author} {\bibfnamefont {D.~A.}\ \bibnamefont {Weitz}},\ }\href@noop {}
  {\bibfield  {journal} {\bibinfo  {journal} {Science}\ }\textbf {\bibinfo
  {volume} {304}},\ \bibinfo {pages} {1301} (\bibinfo {year}
  {2004})}\BibitemShut {NoStop}%
\bibitem [{\citenamefont {Janmey}\ \emph {et~al.}(2007)\citenamefont {Janmey},
  \citenamefont {{McCormick}}, \citenamefont {Rammensee}, \citenamefont
  {Leight}, \citenamefont {Georges},\ and\ \citenamefont
  {MacKintosh}}]{JMRL07}%
  \BibitemOpen
  \bibfield  {author} {\bibinfo {author} {\bibfnamefont {P.~A.}\ \bibnamefont
  {Janmey}}, \bibinfo {author} {\bibfnamefont {M.~E.}\ \bibnamefont
  {{McCormick}}}, \bibinfo {author} {\bibfnamefont {S.}~\bibnamefont
  {Rammensee}}, \bibinfo {author} {\bibfnamefont {J.~L.}\ \bibnamefont
  {Leight}}, \bibinfo {author} {\bibfnamefont {P.~C.}\ \bibnamefont {Georges}},
  \ and\ \bibinfo {author} {\bibfnamefont {F.~C.}\ \bibnamefont {MacKintosh}},\
  }\href@noop {} {\bibfield  {journal} {\bibinfo  {journal} {Nature Mat.}\
  }\textbf {\bibinfo {volume} {6}},\ \bibinfo {pages} {48} (\bibinfo {year}
  {2007})}\BibitemShut {NoStop}%
\bibitem [{\citenamefont {Gerbal}\ \emph {et~al.}(2000)\citenamefont {Gerbal},
  \citenamefont {Chaikin}, \citenamefont {Rabin},\ and\ \citenamefont
  {Prost}}]{GCR00}%
  \BibitemOpen
  \bibfield  {author} {\bibinfo {author} {\bibfnamefont {F.}~\bibnamefont
  {Gerbal}}, \bibinfo {author} {\bibfnamefont {P.}~\bibnamefont {Chaikin}},
  \bibinfo {author} {\bibfnamefont {Y.}~\bibnamefont {Rabin}}, \ and\ \bibinfo
  {author} {\bibfnamefont {J.}~\bibnamefont {Prost}},\ }\href@noop {}
  {\bibfield  {journal} {\bibinfo  {journal} {Biophys. J.}\ }\textbf {\bibinfo
  {volume} {79}},\ \bibinfo {pages} {2259} (\bibinfo {year}
  {2000})}\BibitemShut {NoStop}%
\bibitem [{\citenamefont {Pujol}\ \emph {et~al.}(2012)\citenamefont {Pujol},
  \citenamefont {{du Roure}}, \citenamefont {Fermigier},\ and\ \citenamefont
  {Heuvingh}}]{PRFH12}%
  \BibitemOpen
  \bibfield  {author} {\bibinfo {author} {\bibfnamefont {T.}~\bibnamefont
  {Pujol}}, \bibinfo {author} {\bibfnamefont {O.}~\bibnamefont {{du Roure}}},
  \bibinfo {author} {\bibfnamefont {M.}~\bibnamefont {Fermigier}}, \ and\
  \bibinfo {author} {\bibfnamefont {J.}~\bibnamefont {Heuvingh}},\ }\href@noop
  {} {\bibfield  {journal} {\bibinfo  {journal} {Proc. Natl. Acad. Sci. USA}\
  }\textbf {\bibinfo {volume} {109}},\ \bibinfo {pages} {10364} (\bibinfo
  {year} {2012})}\BibitemShut {NoStop}%
\bibitem [{\citenamefont {Dayel}\ \emph {et~al.}(2009)\citenamefont {Dayel},
  \citenamefont {Akin}, \citenamefont {Landeryou}, \citenamefont {Risca},
  \citenamefont {Mogilner},\ and\ \citenamefont {Mullins}}]{DALR09}%
  \BibitemOpen
  \bibfield  {author} {\bibinfo {author} {\bibfnamefont {M.~J.}\ \bibnamefont
  {Dayel}}, \bibinfo {author} {\bibfnamefont {O.}~\bibnamefont {Akin}},
  \bibinfo {author} {\bibfnamefont {M.}~\bibnamefont {Landeryou}}, \bibinfo
  {author} {\bibfnamefont {V.}~\bibnamefont {Risca}}, \bibinfo {author}
  {\bibfnamefont {A.}~\bibnamefont {Mogilner}}, \ and\ \bibinfo {author}
  {\bibfnamefont {R.~D.}\ \bibnamefont {Mullins}},\ }\href@noop {} {\bibfield
  {journal} {\bibinfo  {journal} {Plos Biol.}\ }\textbf {\bibinfo {volume}
  {7}},\ \bibinfo {pages} {e10000201} (\bibinfo {year} {2009})}\BibitemShut
  {NoStop}%
\bibitem [{\citenamefont {Oelz}\ \emph {et~al.}(2008)\citenamefont {Oelz},
  \citenamefont {Schmeiser},\ and\ \citenamefont {Small}}]{OSS08}%
  \BibitemOpen
  \bibfield  {author} {\bibinfo {author} {\bibfnamefont {D.}~\bibnamefont
  {Oelz}}, \bibinfo {author} {\bibfnamefont {N.}~\bibnamefont {Schmeiser}}, \
  and\ \bibinfo {author} {\bibfnamefont {V.}~\bibnamefont {Small}},\
  }\href@noop {} {\bibfield  {journal} {\bibinfo  {journal} {Cell Adh. Migr.}\
  }\textbf {\bibinfo {volume} {2}},\ \bibinfo {pages} {117} (\bibinfo {year}
  {2008})}\BibitemShut {NoStop}%
\bibitem [{\citenamefont {Kawska}\ \emph {et~al.}(2012)\citenamefont {Kawska},
  \citenamefont {Carvalho}, \citenamefont {Manzi}, \citenamefont
  {Boujemaa-Paterski}, \citenamefont {Blanchoin}, \citenamefont {Martiel},\
  and\ \citenamefont {Sykes}}]{Kawska_PNAS_2012}%
  \BibitemOpen
  \bibfield  {author} {\bibinfo {author} {\bibfnamefont {A.}~\bibnamefont
  {Kawska}}, \bibinfo {author} {\bibfnamefont {K.}~\bibnamefont {Carvalho}},
  \bibinfo {author} {\bibfnamefont {J.}~\bibnamefont {Manzi}}, \bibinfo
  {author} {\bibfnamefont {R.}~\bibnamefont {Boujemaa-Paterski}}, \bibinfo
  {author} {\bibfnamefont {L.}~\bibnamefont {Blanchoin}}, \bibinfo {author}
  {\bibfnamefont {J.-L.}\ \bibnamefont {Martiel}}, \ and\ \bibinfo {author}
  {\bibfnamefont {C.}~\bibnamefont {Sykes}},\ }\href@noop {} {\bibfield
  {journal} {\bibinfo  {journal} {Proc. Natl. Acad. Sci. USA}\ }\textbf
  {\bibinfo {volume} {109}},\ \bibinfo {pages} {14440} (\bibinfo {year}
  {2012})}\BibitemShut {NoStop}%
\bibitem [{\citenamefont {Sekimoto}\ \emph {et~al.}(2004)\citenamefont
  {Sekimoto}, \citenamefont {Prost}, \citenamefont {J\"ulicher}, \citenamefont
  {Boukellal},\ and\ \citenamefont {Bernheim-Grosswasser}}]{SPJ04}%
  \BibitemOpen
  \bibfield  {author} {\bibinfo {author} {\bibfnamefont {K.}~\bibnamefont
  {Sekimoto}}, \bibinfo {author} {\bibfnamefont {J.}~\bibnamefont {Prost}},
  \bibinfo {author} {\bibfnamefont {F.}~\bibnamefont {J\"ulicher}}, \bibinfo
  {author} {\bibfnamefont {H.}~\bibnamefont {Boukellal}}, \ and\ \bibinfo
  {author} {\bibfnamefont {A.}~\bibnamefont {Bernheim-Grosswasser}},\
  }\href@noop {} {\bibfield  {journal} {\bibinfo  {journal} {Eur. Phys. J. E}\
  }\textbf {\bibinfo {volume} {13}},\ \bibinfo {pages} {247} (\bibinfo {year}
  {2004})}\BibitemShut {NoStop}%
\bibitem [{\citenamefont {Blanchoin}\ \emph {et~al.}(2014)\citenamefont
  {Blanchoin}, \citenamefont {Boujemaa-Paterski}, \citenamefont {Sykes},\ and\
  \citenamefont {Plastino}}]{Blanchoin_PhysRev_2014}%
  \BibitemOpen
  \bibfield  {author} {\bibinfo {author} {\bibfnamefont {L.}~\bibnamefont
  {Blanchoin}}, \bibinfo {author} {\bibfnamefont {R.}~\bibnamefont
  {Boujemaa-Paterski}}, \bibinfo {author} {\bibfnamefont {C.}~\bibnamefont
  {Sykes}}, \ and\ \bibinfo {author} {\bibfnamefont {J.}~\bibnamefont
  {Plastino}},\ }\href@noop {} {\bibfield  {journal} {\bibinfo  {journal}
  {Physiol. Rev.}\ }\textbf {\bibinfo {volume} {94}},\ \bibinfo {pages} {235}
  (\bibinfo {year} {2014})}\BibitemShut {NoStop}%
\bibitem [{\citenamefont {Mogilner}\ and\ \citenamefont {Oster}(2003)}]{MoO03}%
  \BibitemOpen
  \bibfield  {author} {\bibinfo {author} {\bibfnamefont {A.}~\bibnamefont
  {Mogilner}}\ and\ \bibinfo {author} {\bibfnamefont {G.}~\bibnamefont
  {Oster}},\ }\href@noop {} {\bibfield  {journal} {\bibinfo  {journal}
  {Biophys. J.}\ }\textbf {\bibinfo {volume} {84}},\ \bibinfo {pages} {1591}
  (\bibinfo {year} {2003})}\BibitemShut {NoStop}%
\bibitem [{\citenamefont {Lee}\ \emph {et~al.}(2005)\citenamefont {Lee},
  \citenamefont {Lee},\ and\ \citenamefont {Kardar}}]{LLK05}%
  \BibitemOpen
  \bibfield  {author} {\bibinfo {author} {\bibfnamefont {A.}~\bibnamefont
  {Lee}}, \bibinfo {author} {\bibfnamefont {H.~Y.}\ \bibnamefont {Lee}}, \ and\
  \bibinfo {author} {\bibfnamefont {M.}~\bibnamefont {Kardar}},\ }\href@noop {}
  {\bibfield  {journal} {\bibinfo  {journal} {Phys. Rev. Lett.}\ }\textbf
  {\bibinfo {volume} {95}},\ \bibinfo {pages} {138101} (\bibinfo {year}
  {2005})}\BibitemShut {NoStop}%
\bibitem [{\citenamefont {John}\ \emph {et~al.}(2008)\citenamefont {John},
  \citenamefont {Peyla}, \citenamefont {Kassner}, \citenamefont {Prost},\ and\
  \citenamefont {Misbah}}]{JPK08}%
  \BibitemOpen
  \bibfield  {author} {\bibinfo {author} {\bibfnamefont {K.}~\bibnamefont
  {John}}, \bibinfo {author} {\bibfnamefont {P.}~\bibnamefont {Peyla}},
  \bibinfo {author} {\bibfnamefont {K.}~\bibnamefont {Kassner}}, \bibinfo
  {author} {\bibfnamefont {J.}~\bibnamefont {Prost}}, \ and\ \bibinfo {author}
  {\bibfnamefont {C.}~\bibnamefont {Misbah}},\ }\href@noop {} {\bibfield
  {journal} {\bibinfo  {journal} {Phys. Rev. Lett.}\ }\textbf {\bibinfo
  {volume} {100}},\ \bibinfo {pages} {068101} (\bibinfo {year}
  {2008})}\BibitemShut {NoStop}%
\bibitem [{\citenamefont {John}\ \emph {et~al.}(2013)\citenamefont {John},
  \citenamefont {Caillerie}, \citenamefont {Peyla}, \citenamefont {Raoult},\
  and\ \citenamefont {Misbah}}]{John_PhysRevE2013}%
  \BibitemOpen
  \bibfield  {author} {\bibinfo {author} {\bibfnamefont {K.}~\bibnamefont
  {John}}, \bibinfo {author} {\bibfnamefont {D.}~\bibnamefont {Caillerie}},
  \bibinfo {author} {\bibfnamefont {P.}~\bibnamefont {Peyla}}, \bibinfo
  {author} {\bibfnamefont {A.}~\bibnamefont {Raoult}}, \ and\ \bibinfo {author}
  {\bibfnamefont {C.}~\bibnamefont {Misbah}},\ }\href@noop {} {\bibfield
  {journal} {\bibinfo  {journal} {Phys. Rev. E}\ }\textbf {\bibinfo {volume}
  {87}},\ \bibinfo {pages} {042721} (\bibinfo {year} {2013})}\BibitemShut
  {NoStop}%
\bibitem [{\citenamefont {Doedel}\ \emph {et~al.}(1998)\citenamefont {Doedel},
  \citenamefont {Champneys}, \citenamefont {Fairgrieve}, \citenamefont
  {Kuznetsov}, \citenamefont {B.},\ and\ \citenamefont {Wang}}]{AUTO97}%
  \BibitemOpen
  \bibfield  {author} {\bibinfo {author} {\bibfnamefont {E.~J.}\ \bibnamefont
  {Doedel}}, \bibinfo {author} {\bibfnamefont {A.~R.}\ \bibnamefont
  {Champneys}}, \bibinfo {author} {\bibfnamefont {T.~F.}\ \bibnamefont
  {Fairgrieve}}, \bibinfo {author} {\bibfnamefont {Y.~A.}\ \bibnamefont
  {Kuznetsov}}, \bibinfo {author} {\bibfnamefont {S.}~\bibnamefont {B.}}, \
  and\ \bibinfo {author} {\bibfnamefont {X.~J.}\ \bibnamefont {Wang}},\
  }\href@noop {} {\bibfield  {journal} {\bibinfo  {journal} {AUTO97:
  Continuation and bifurcation software for ordinary differential equations}\ }
  (\bibinfo {year} {1998})}\BibitemShut {NoStop}%
\bibitem [{\citenamefont {Pollard}\ \emph {et~al.}(2000)\citenamefont
  {Pollard}, \citenamefont {Blanchoin},\ and\ \citenamefont {Mullins}}]{PBM00}%
  \BibitemOpen
  \bibfield  {author} {\bibinfo {author} {\bibfnamefont {T.~D.}\ \bibnamefont
  {Pollard}}, \bibinfo {author} {\bibfnamefont {L.}~\bibnamefont {Blanchoin}},
  \ and\ \bibinfo {author} {\bibfnamefont {R.~D.}\ \bibnamefont {Mullins}},\
  }\href@noop {} {\bibfield  {journal} {\bibinfo  {journal} {Ann. Rev. Biophys.
  Biomol. Struct.}\ }\textbf {\bibinfo {volume} {29}},\ \bibinfo {pages} {545}
  (\bibinfo {year} {2000})}\BibitemShut {NoStop}%
\bibitem [{\citenamefont {Kruse}\ \emph {et~al.}(2004)\citenamefont {Kruse},
  \citenamefont {Joanny}, \citenamefont {J\"ulicher}, \citenamefont {Prost},\
  and\ \citenamefont {Sekimoto}}]{KJJP04}%
  \BibitemOpen
  \bibfield  {author} {\bibinfo {author} {\bibfnamefont {K.}~\bibnamefont
  {Kruse}}, \bibinfo {author} {\bibfnamefont {J.~F.}\ \bibnamefont {Joanny}},
  \bibinfo {author} {\bibfnamefont {F.}~\bibnamefont {J\"ulicher}}, \bibinfo
  {author} {\bibfnamefont {J.}~\bibnamefont {Prost}}, \ and\ \bibinfo {author}
  {\bibfnamefont {K.}~\bibnamefont {Sekimoto}},\ }\href@noop {} {\bibfield
  {journal} {\bibinfo  {journal} {Phys. Rev. Lett.}\ }\textbf {\bibinfo
  {volume} {92}},\ \bibinfo {pages} {078101} (\bibinfo {year}
  {2004})}\BibitemShut {NoStop}%
\end{thebibliography}

%

\end{document}